\documentclass[a4paper,12pt]{article}

\usepackage{amsmath,amssymb}
\usepackage[dvips]{graphicx}
\begin{document}
\renewcommand{\thefootnote}{\fnsymbol{footnote}}

\begin{titlepage}
\vspace*{5mm}
\hfill
{\hfill \begin{flushright} 
YITP-06-17 \\ 
hep-th/0604194
\end{flushright}
  }

\vspace*{10mm}

\begin{center}
{\LARGE {\LARGE  
One-loop unitarity of scalar field theories on Poincar\'{e} invariant commutative nonassociative spacetimes}}
\vspace*{15mm}

 {\large Yuya Sasai}
\footnote{ e-mail: sasai@yukawa.kyoto-u.ac.jp}
 {\large and~ Naoki Sasakura}
\footnote{ e-mail: sasakura@yukawa.kyoto-u.ac.jp}

\vspace*{7mm}
{\large {\it 
Yukawa Institute for Theoretical Physics, Kyoto University, \\ 
 Kyoto 606-8502, Japan 
}} \\

\end{center}

\vspace*{1.5cm}

\begin{abstract}
We study scalar field theories on Poincar\'{e} invariant commutative nonassociative spacetimes. 
We compute the one-loop self-energy diagrams in the ordinary path integral quantization scheme with
Feynman's prescription, and find that the Cutkosky rule is satisfied.
This property is in contrast with that of noncommutative field theory, 
since it is known that noncommutative field theory with
space/time noncommutativity violates unitarity in the above standard scheme, 
and the quantization procedure will necessarily become complicated to obtain a sensible Poincar\'e invariant
noncommutative field theory.
We point out a peculiar feature of the non-locality in our nonassociative field theories, which may explain the property of the unitarity distinct from noncommutative field theories. 
Thus commutative nonassociative field theories seem to contain physically  
interesting field theories on deformed spacetimes. 
\end{abstract}

\end{titlepage}
\newpage

\section{Introduction}
\renewcommand{\thefootnote}{\arabic{footnote}}
\setcounter{footnote}{0}
Noncommutative spacetime is spacetime with noncommutative spacetime coordinates 
\cite{snyder}-\cite{Doplicher:1994tu}. 
Since there appears uncertainty in non-commuting directions, 
noncommutative spacetime can be considered 
as an interesting candidate for new notion of quantum spacetime. 
Field theories on noncommutative spacetime are obtained from replacing 
commutative $C^{\ast}$-algebras of functions with noncommutative algebras \cite{connes}. 

It is known that noncommutative field theories are related to string theory and quantum gravity. In string theory, for example, noncommutative field theory is an effective theory, which is obtained in the 
$\alpha' \to 0$ limit of the open string theory with constant background $B_{\mu\nu}$ field \cite{Seiberg:1999vs}. In quantum gravity, for example, the effective dynamics of quantum particles coupled to three-dimensional quantum
gravity can be expressed in terms of an effective noncommutative field theory which respects the principles of doubly special relativity \cite{Freidel:2005bb}.

But some noncommutative field theories do not respect the principles of special relativity or quantum mechanics. 
The simplest example is the spacetime with the noncommutativity $[x^{\mu}, x^{\nu}]=i\theta^{\mu\nu}$, 
where $\theta^{\mu\nu}$ is a constant antisymmetric tensor. The field theory on the noncommutative spacetime does not have the ordinary Lorentz symmetry except in two-dimensions, 
since $\theta^{\mu\nu}$ is not Lorentz invariant\footnote{In two dimensions, $\theta^{\mu\nu}$ is proportional to $\epsilon^{\mu\nu}$ and is Lorentz invariant.}, while recent developments show that noncommutative field theory has twisted Poincar\'{e} 
symmetry \cite{Chaichian:2004za, Chaichian:2004yh}. It is not unitary
in the standard path integral quantization procedure with Feynman's prescription, 
when it has space/time noncommutativity \cite{Gomis:2000zz}-\cite{Chu:2002fe}. 
In \cite{Gomis:2000zz}, this was shown by computing explicitly the one-loop amplitudes 
and showing the violation of the Cutkosky rule\footnote{In \cite{Bahns:2002vm, Rim:2002if},
however, they defined a unitary S-matrix of space/time noncommutative field theory 
with proper time-ordering.
The amplitudes in their schemes are different from those in the standard path integral
scheme. }. 
Also it is known that acausal effects occur, when it has space/time noncommutativity \cite{Seiberg:2000gc}. 
Moreover, it has an unusual behavior, called UV-IR mixing, that 
the ultra-violet divergences appear in the limit of vanishing external momenta 
\cite{Filk:1996dm}-\cite{Hayakawa:1999zf}.

Is it possible to construct field theories on deformed spacetime which preserve both Lorentz symmetry and 
unitarity? Since Lorentz symmetry mixes space and time directions, the known facts in the preceding paragraph 
suggest that it becomes necessarily complicated for noncommutative spacetime. 
In this paper, we pursue the possibility 
of nonassociative spacetime.  In fact, nonassociativity is known to appear in open string theory with non-constant background $B_{\mu\nu}$ field \cite{Cornalba:2001sm}. It was also argued that 
the algebra of closed string field theory should be commutative nonassociative \cite{Witten:1985cc}. 
There are also other discussions on nonassociative theory \cite{Ootsuka:2005ay, Majid:2005kp}. 
Especially in \cite{deMedeiros:2004wb}, they discussed commutative nonassociative gauge theory 
with Lorentz symmetry. 

This paper is organized as follows.  
In the following section we study scalar $\phi^3$ field theory 
obtained from the commutative nonassociative product,
\begin{equation*}
\phi(x)\ast \phi(x)=e^{-\alpha(\partial_{a}+\partial_{b})^2}\phi(x+a)\phi(x+b)|_{a=b=0}, 
\end{equation*}
where $\alpha$ is a constant nonassociative deformation parameter.
This product is obviously Poincar\'{e} invariant. 
Since this product contains an infinite number of space-time derivatives, unitarity 
seems to be a non-trivial issue. 
We find that this field theory satisfies the Cutkosky rule for the one-loop self-energy diagram.

In Section \ref{sec:p4} we replace the above commutative nonassociative product to avoid a divergence in the amplitude.  
In fact the real part of the one-loop amplitude diverges exponentially in Minkowski spacetime,
when we adopt the above product. 
This divergence is irrelevant to the discussions about the one-loop unitarity in Section \ref{sec:p2}, 
but may harm the significance
of the field theory based on the above product.  
Thus we change the square of momenta on the exponential to the forth power, and study the 
field theory based on the new product. We check the one-loop Cutkosky rule 
to some orders of $\alpha$, and find the unitarity holds also in this case.

In Section \ref{sec:gen} we discuss scalar field theories obtained from 
Poincar\'{e} invariant commutative associative algebras, and find that the couplings become constant in general. 
This shows that non-trivial behaviors of scalar field theories 
can appear only when commutativity or associativity of algebras is lost. 

The final section is devoted to discussions and comments.
We make an observation concerning the reason why our nonassociative field theories 
satisfy the unitarity relation from the viewpoint of a qualitative difference 
in non-locality between noncommutative field theories and our commutative nonassociative field theories. 

\section{Nonassociative $\phi^3$ theory: quadratic case} \label{sec:p2}
Noncommutative field theories can be constructed from replacing the usual multiplication of fields 
in the Lagrangian with Moyal product,
\begin{equation}
\phi(x)\ast \phi(x)=e^{\frac{i}{2}\theta^{\mu\nu}\frac{\partial}{\partial a^{\mu}}\frac{\partial}{\partial b^{\nu}}}\phi(x+a)\phi(x+b)|_{a=b=0}. \label{eq:moyal}
\end{equation}
Then the spacetime noncommutativity is given by
\begin{equation}
[x^{\mu},x^{\nu}]=x^{\mu}\ast x^{\nu}-x^{\nu}\ast x^{\mu}=i\theta^{\mu\nu}.
\end{equation}
This spacetime noncommutativity breaks Lorentz symmetry except in two dimensions. 
Moreover field theory with space/time noncommutativity ($\theta^{0i}\neq 0$) is not unitary
in the standard path integral scheme
\cite{Gomis:2000zz}-\cite{Chu:2002fe}.

The purpose of this paper is to pursue unitary field theories 
on deformed spacetimes with  Lorentz symmetry.
As candidates, we are going to construct scalar field theories on Poincar\'{e} invariant commutative nonassociative 
spacetimes and check the one-loop unitarity.

Let us first define the following commutative nonassociative star product,
\begin{equation}
\phi(x)\ast \phi(x)=e^{-\alpha(\partial_{a}+\partial_{b})^2}\phi(x+a)\phi(x+b)|_{a=b=0}, \label{eq:p2nonas}
\end{equation}
where $\alpha$ is a constant nonassociative deformation parameter. 
This parameter is taken to be real for the tree level unitarity to hold. 

For the plane waves, the product (\ref{eq:p2nonas}) becomes
\begin{equation}
e^{ipx}\ast e^{iqx}=e^{\alpha (p+q)^2}e^{i(p+q)x}. \label{eq:momentumrep}
\end{equation}
This is obviously Poincar\'{e} invariant, since it preserves the Lorentz symmetry and the momentum conservation.
One can easily check that this product is commutative but nonassociative,
\begin{align}
e^{ip_1x}\ast e^{ip_2x}&=e^{ip_2x}\ast e^{ip_1x}, \notag \\
(e^{ip_1x}\ast e^{ip_2x})\ast e^{ip_3x}&\neq e^{ip_1x}\ast (e^{ip_2x}\ast e^{ip_3x}).
\end{align}
Field theories based on the product\footnote{
We adopt the above product (\ref{eq:momentumrep}),
 because this is the simplest non-trivial choice if we impose the product 
to have the exponential form and Poincar\'{e} invariance. You may hit upon another choice,
\[e^{ipx}\ast e^{iqx}=e^{\alpha p\cdot q}e^{i(p+q)x}. \]
But this product is commutative associative, and field theory based on it is trivial in the sense
discussed in Section \ref{sec:gen}. In fact, one can multiply the right hand side of (\ref{eq:momentumrep}) 
by any  commutative associative factor
with no change of field theory (See Section \ref{sec:gen} for more details.). 
Therefore the expression (\ref{eq:momentumrep}) is essentially the general case with the form of the exponential of a
quadratic function of momenta. The peculiar choice (\ref{eq:momentumrep}) is taken for the simplest choice 
of the normalization of the scalar field.} 
will have features quite different from the noncommutative field theory based on the Moyal product.

Let us construct $\phi^3$ scalar field theory based on the above commutative nonassociative product.
The action is defined by
\begin{equation}
\label{phithreeaction}
S=\int d^Dx \bigg[\frac{1}{2}\partial_{\mu}\phi(x)\ast \partial^{\mu}\phi(x)-\frac{1}{2}m^2\phi(x)\ast \phi(x)-\frac{g}{3!}\phi(x)\ast (\phi(x)\ast \phi(x))\bigg].
\end{equation}
The term $(\phi \ast \phi)\ast \phi$ is not necessary, because $(\phi \ast \phi)\ast \phi =\phi \ast (\phi\ast \phi)$
holds from the commutativity. 

\begin{figure}
\begin{center}
\includegraphics[scale=0.5]{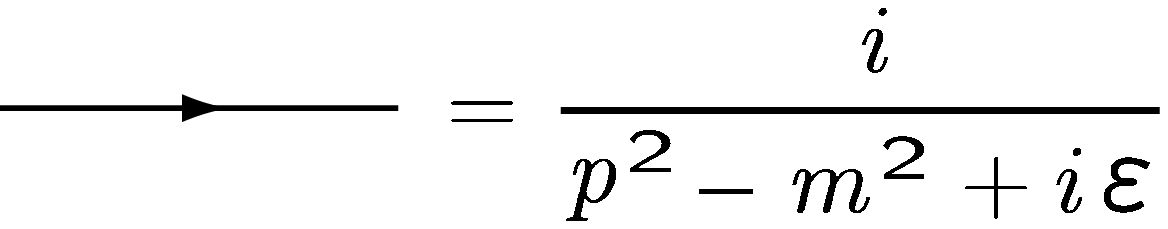}
\end{center}
\caption{Propagator}
\label{fig:prop}
\end{figure}

\begin{figure}
\begin{center}
\includegraphics[scale=0.5]{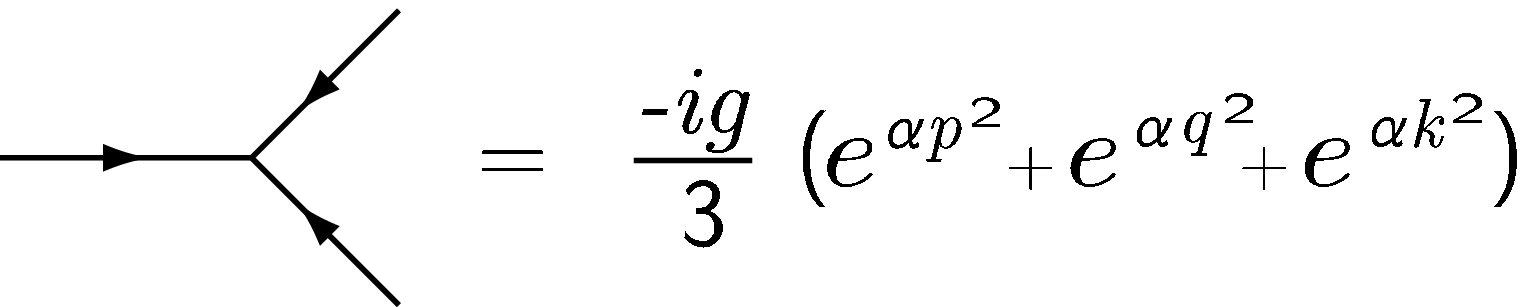}
\end{center}
\caption{Three-point vertex}
\label{fig:3pointver}
\end{figure}

In this paper, we employ the standard path integral quantization procedure for the action
(\ref{phithreeaction}). 
To find the Feynman rules, let us consider the Fourier transform of the field $\phi(x)$,
\begin{equation}
\label{eq:momphi}
\phi(x)=\int_p \tilde{\phi}(p)\ e^{ipx},
\end{equation}
where $\int_p=\int d^Dp/(2\pi )^D$. The scalar field $\phi(x)$ is assumed to be real, and
therefore $\tilde\phi(p)^\ast=\tilde\phi(-p)$.
Substituting (\ref{eq:momphi}) into the action, we obtain
\begin{align}
S&=\int_p\frac{1}{2}(p^2-m^2)\tilde{\phi}(p)\tilde{\phi}(-p) \notag \\
&\ \ \ \ -\frac{g}{3!}\int_p\int_q\int_k e^{\alpha (q+k)^2}e^{\alpha (p+(q+k))^2}\tilde{\phi}(p)\tilde{\phi}(q)\tilde{\phi}(k)(2\pi)^D\delta^D(p+q+k) \notag \\
&=\int_p\frac{1}{2}(p^2-m^2)\tilde{\phi}(p)\tilde{\phi}(-p) 
-\frac{g}{3!}\int_p\int_q e^{\alpha p^2}\tilde{\phi}(p)\tilde{\phi}(q)\tilde{\phi}(-p-q).
\label{eq:momact}
\end{align}
One can read Feynman rules from (\ref{eq:momact}).
The Feynman rule for the propagator is given by the usual one as in Figure \ref{fig:prop}. 
Averaging over the orderings of the legs, 
the Feynman rule of the three-point vertex is given by Figure \ref{fig:3pointver}.
Formal discussions on the quantization of the system respecting the structure of the star product 
will be given below.

Generating functional is defined by
\begin{align}
W[J]&=\int [d\phi]\exp{\bigg\{i\int d^Dx\bigg(\frac{1}{2}\partial_{\mu}\phi\ast \partial^{\mu}\phi-\frac{1}{2}(m^2-i\epsilon )\phi \ast \phi
+J\ast \phi+\mathcal{L}_I[\phi ]\bigg)\bigg\}} \notag \\
&=\exp{\bigg\{\int d^Dx\bigg(i\mathcal{L}_{I}\bigg[\frac{\delta}{i\delta J}\bigg]\bigg) \bigg\}}W_0[J], 
\label{eq:genfun}
\end{align}
where
\begin{align}
W_0[J]&=\int [d\phi]\exp{\bigg\{i\int d^Dx\bigg(\frac{1}{2}\partial_{\mu}\phi\ast \partial^{\mu}\phi-\frac{1}{2}(m^2-i\epsilon)\phi\ast \phi+J\ast \phi\bigg)\bigg\}} \notag \\
&=\int [d\phi]\exp{\bigg\{i\int d^Dx\bigg(\frac{1}{2}\phi \ast (-\partial ^2-m^2+i\epsilon)\phi+J\ast \phi\bigg)\bigg\}}, \\
\mathcal{L}_{I}[\phi]&=-\frac{g}{3!}\phi\ast (\phi\ast \phi).
\end{align}
We have inserted a factor $\epsilon \sim +0$ to make the path integral convergent as usual.
In deriving the second line of (\ref{eq:genfun}), we have used that $\delta/(i\delta J(x))$
can be replaced with $\phi(x)$. This can be shown as 
\begin{align}
&\frac{\delta }{i\delta J(x')}\exp{\bigg(i\int d^DxJ(x)\ast\phi(x)\bigg)} \notag \\
&=\int d^Dx \delta(x'-x) \ast_x \phi(x)\exp{\bigg(i\int d^DxJ(x)\ast\phi(x)\bigg)} \notag \\
&=\int d^Dx \int_p \int_k e^{ip(x'-x)}\ast_x e^{ikx}\tilde{\phi}(k)\exp{\bigg(i\int d^DxJ(x)\ast\phi(x)\bigg)} \notag \\
&=\int d^Dx \int_p \int_k e^{\alpha (-p+k)^2}e^{i(-p+k)x}e^{ipx'}\tilde{\phi}(k)\exp{\bigg(i\int d^DxJ(x)\ast\phi(x)\bigg)} \notag \\
&=\int_p \int_ke^{\alpha (p-k)^2}(2\pi)^D \delta^D(-p+k)e^{ipx'}\tilde{\phi}(k)\exp{\bigg(i\int d^DxJ(x)\ast\phi(x)\bigg)} \notag \\
&=\int_p\tilde{\phi}(p)e^{ipx'}\exp{\bigg(i\int d^DxJ(x)\ast\phi(x)\bigg)} \notag \\
&=\phi(x')\exp{\bigg(i\int d^DxJ(x)\ast\phi(x)\bigg)}.
\end{align}

We now change the integration variable from $\phi(x)$ to $\phi'(x)$ defined by
\begin{equation}
\phi(x)=\phi_c(x)+\phi'(x),
\end{equation}
where $\phi_c(x)$ is a solution to the classical free field equation, 
\begin{equation}
(-\partial^2-m^2+i\epsilon)\phi_c(x)=-J(x).
\end{equation}
The solution is given by
\begin{equation}
\phi_c(x)=-\int d^Dy\Delta_F(x-y)J(y),
\end{equation}
where 
\begin{equation}
\Delta_F(x-y)=\int \frac{d^Dk}{(2\pi )^D}e^{-ik(x-y)}\frac{1}{k^2-m^2+i\epsilon}.
\end{equation}
Then we get 
\begin{equation}
W_0[J]=N\exp{\bigg\{-\frac{i}{2}\int d^Dx\int d^DyJ(x)\ast_x \Delta_F(x-y)J(y)\bigg\}},
\end{equation}
where
\begin{equation}
N=\int [d\phi']\exp{\bigg\{i\int d^Dx\frac{1}{2}\phi'\ast (-\partial^2-m^2+i\epsilon)\phi'\bigg\}}.
\end{equation}
Note that the factor $N$ is independent of the current $J(x)$. 

The connected two point function of free field theory is given by
\begin{align}
\langle 0|T\phi(x)\phi(y)|0\rangle &=\frac{\delta^2\ln W_0[J]}{i\delta J(x)i\delta J(y)}\bigg|_{J=0} \notag \\
&=i\int d^Dx'\int d^Dy' \delta^D(x-x')\ast_{x'}\Delta_F(x'-y')\delta^D(y-y') \notag \\
&=i\int \frac{d^Dp}{(2\pi )^D}\int \frac{d^Dk}{(2\pi )^D}\delta^D(p-k)e^{\alpha(p-k)^2}e^{-ipx}e^{iky}\frac{1}{k^2-m^2+i\epsilon} \notag \\
&=\int \frac{d^Dp}{(2\pi)^D}\frac{i}{p^2-m^2+i\epsilon}e^{-ip(x-y)}.
\end{align}
Thus the propagator is the usual one,
\begin{equation}
\frac{i}{p^2-m^2+i\epsilon}.
\end{equation}

The connected three point function is given by
\begin{equation}
\langle 0|T\phi(x)\phi(y)\phi(z)|0\rangle =\frac{\delta^3\ln W[J]}{i\delta J(x)i\delta J(y)i\delta J(z)}\bigg|_{J=0}.
\end{equation}
From the tree level contribution, we obtain the Feynman rule for the three-point vertex as 
\begin{equation}
\frac{-ig}{3}(e^{\alpha p^2}+e^{\alpha q^2}+e^{\alpha k^2}),
\end{equation}
where $p, q, k$ are the external momenta.

\begin{figure}
\begin{center}
\includegraphics[scale=.5]{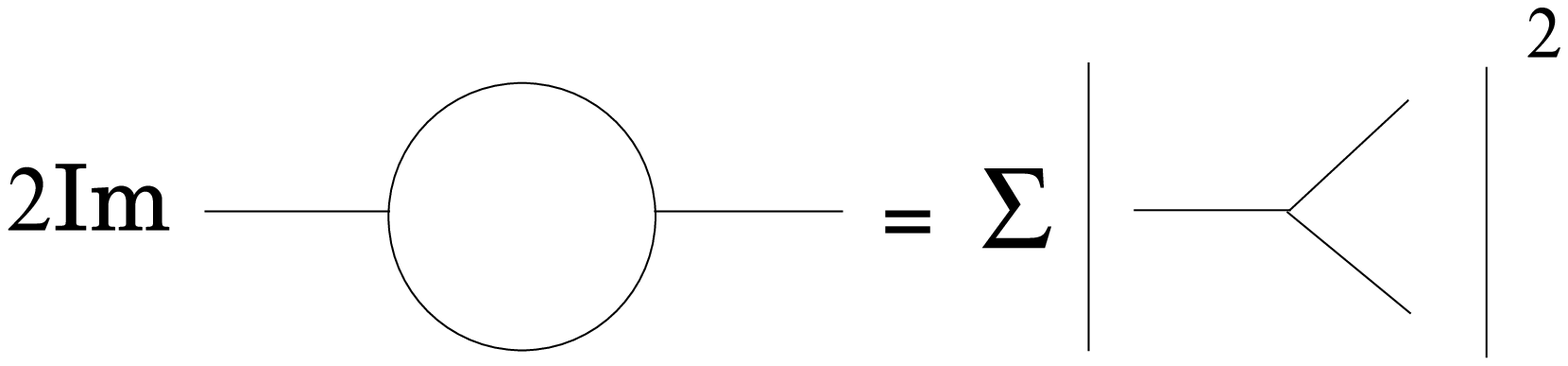}
\end{center}
\caption{One-loop Cutkosky rule in $\phi^3$ theory.}
\label{fig:Cutkosky}
\end{figure}

A unitary theory will satisfy the Cutkosky rule,
\begin{equation}
2\mathrm{Im}\mathcal{M}_{ab}=\sum_n\mathcal{M}_{an}\mathcal{M}_{nb} , \label{eq:Cutkosky}
\end{equation}
where $\mathcal{M}_{ab}$ is the transition matrix element between states $a$ and $b$. Using the Feynman rules (Figures \ref{fig:prop}, \ref{fig:3pointver}), we will check the Cutkosky rule for the one-loop self-energy diagram of 
the commutative nonassociative $\phi^3$ theory.
The rule is diagrammatically given by Figure~\ref{fig:Cutkosky}.

Let us compute the one-loop amplitude $\mathcal{M}$ in Figure \ref{fig:Cutkosky}. 
The amplitude is given by 
\begin{equation}
\begin{split}
i\mathcal{M}=\bigg(\frac{-ig}{3} \bigg)^2\frac{1}{2}&\int \frac{d^Dq}{(2\pi )^D}
(e^{\alpha p^2}+e^{\alpha q^2}+e^{\alpha (p+q)^2})^2 \\
&\cdot \frac{i}{q^2-m^2+i\epsilon }\frac{i}{(p+q)^2-m^2+i\epsilon }.
\label{eq:mrepexp1}
\end{split}
\end{equation}
If $\alpha$ is non-zero, the momentum integration diverges exponentially in Minkowski spacetime. 
This can be cured by
changing the star product, which will be discussed in the following section.
In this section, however, we will stick to the star product (\ref{eq:p2nonas})
and compute the amplitude by analytic continuation. 
One reason for this is that what diverges is actually the real part of the 
one-loop amplitude, which is irrelevant to the one-loop unitarity. 
Another reason is that we can obtain a definite conclusion if we adopt the star
product (\ref{eq:p2nonas}), because the imaginary part of the one-loop amplitude can be computed exactly in $D=3$. 

Let us assume $\alpha>0$ and carry out the Wick rotation of the amplitude. 
Then, after combining the denominators by using the Feynman parameters, we obtain
\begin{equation}
\label{ampfirst}
\begin{split}
\mathcal{M}=\frac{g^2}{18}\int \frac{d^Dq_E}{(2\pi )^D}&
(e^{-\alpha p_E^2}+e^{-\alpha q_E^2}+e^{-\alpha (p_E+q_E)^2})^2 \\
&\int_0^1dx\frac{1}{((q_E+p_E(1-x))^2+p_E^2x(1-x)+m^2-i\epsilon )^2},
\end{split}
\end{equation}
where $q_E,p_E$ are Euclidean momenta\footnote{\label{foot:rot}
The signature in the Minkowski spacetime is taken as $p^2=(p_0)^2-\sum_{i=1}^{D-1} (p_i)^2$. After Wick rotation
to the Euclidean space,
$p^2=-(p_0)^2-\sum_{i=1}^{D-1}(p_i)^2=-(p_E)^2$. }.
After carrying out the momentum shift $q_E+p_E(1-x)\to q_E$, we obtain
\begin{align}
\mathcal{M}=
\frac{g^2}{18}\int \frac{d^Dq_E}{(2\pi )^D}\int_0^1dx&(e^{-2\alpha p_E^2} \label{eq:unitaryfir} \\
&+e^{-2\alpha (q_E-p_E(1-x))^2} \label{eq:unitarysec}  \\
&+e^{-2\alpha (q_E+p_Ex)^2} \label{eq:unitarythi} \\ 
&+2e^{-\alpha p_E^2}e^{-\alpha (q_E-p_E(1-x))^2} \label{eq:unitaryfor} \\
&+2e^{-\alpha (q_E-p_E(1-x))^2}e^{-\alpha (q_E+p_Ex)^2} \label{eq:unitaryfif}  \\
&+2e^{-\alpha p_E^2}e^{-\alpha (q_E+p_Ex)^2}) \label{eq:unitarysix}  \\ 
&\cdot \frac{1}{(q_E^2+p_E^2x(1-x)+m^2-i\epsilon )^2}. \notag
\end{align}

From now on, let us assume $D=3$.
Let us first study the contribution of the first term (\ref{eq:unitaryfir}).
After parameterizing the three-momentum $q_E$ with 
the radial and angular variables and integrating over the latter, we obtain
\begin{equation}
\mathcal{M}_1=e^{2\alpha p^2}\frac{g^2}{36\pi ^2}\int_0^{\infty }dq_E\int_0^1dx \frac{q_E^2}{(q_E+A_{\epsilon})^2(q_E-A_{\epsilon})^2},
\label{eq:Mfirstterm} 
\end{equation}
where 
\[A_{\epsilon}=\sqrt{p^2 x(1-x)-m^2+i\epsilon}.\]
Here we have carried out the replacement 
$(p_E)^2=-p^2$ to go back to Minkowski spacetime.
In the region $p^2 x(1-x)-m^2<0$, $A_\epsilon$ approaches a pure imaginary value in the 
$\epsilon\rightarrow+0$ limit, and the integrand in 
(\ref{eq:Mfirstterm}) is obviously real. However, when $p^2 x(1-x)-m^2>0$, $A_\epsilon$ approaches a positive real
value in the $\epsilon\rightarrow+0$ limit, and a careful treatment is required for the integration over $q_E$.   
In $x$, this range is expressed as 
\begin{equation}
1/2-\gamma\leq x \leq 1/2+\gamma,
\end{equation}
where $\gamma =\sqrt{\frac{p^2-4m^2}{4p^2}}~(>0)$. Note that $p^2>4m^2$ must be satisfied for the imaginary
part of the amplitude to exist. In this range, 
\begin{equation}
A_{\epsilon} \sim \sqrt{p^2x(1-x)-m^2}+i\epsilon
\equiv A+i\epsilon.
\end{equation}

\begin{figure}
\begin{center}
\includegraphics[scale=.7]{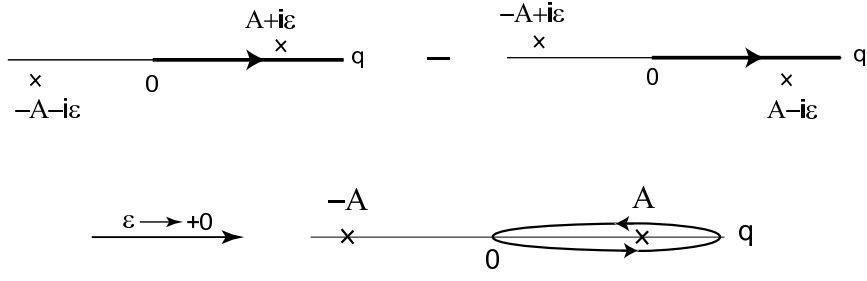}
\end{center}
\caption{The contour of momentum $q_E$}
\label{fig:contour1}
\end{figure}

The imaginary part of $\mathcal{M}_1$ is given by $(\mathcal{M}_1-\mathcal{M}_1^{\ast})/2i$. As can be 
understood from Figure \ref{fig:contour1}, the imaginary part of the amplitude is given by
the contour integration of $q_E$ around the pole $q_E=A$. 
Carrying out the contour integration and the integration over $x$, we obtain
\begin{align}
2\mathrm{Im}\mathcal{M}_1 
&=2e^{2\alpha p^2}\frac{g^2}{36\pi ^2}\int_{1/2-\gamma }^{1/2+\gamma }dx\frac{\pi }{4\sqrt{p^2x(1-x)-m^2}} \notag  \\
&=\frac{g^2e^{2\alpha p^2}}{72\sqrt{p^2}},
\label{eq:ans1p2}
\end{align}
where the branch of positive values is taken for $\sqrt{p^2}$. 

The integration over $q_E$ for the second term (\ref{eq:unitarysec}) can be computed similarly. 
After integrating over the angular variables of $q_E$, we get
\begin{equation*}
\begin{split}
\mathcal{M}_2=\frac{g^2}{72\pi ^2}\int_0^{\infty }dq_E\int _0^1dxe^{-2\alpha p_E^2(1-x)^2}&e^{-2\alpha q_E^2}q_E^2
\frac{e^{4\alpha (1-x)p_Eq_E}-e^{-4\alpha (1-x)p_Eq_E}}{4\alpha (1-x)p_Eq_E} \\
&\cdot \frac{1}{(q_E^2+p_E^2x(1-x)+m^2-i\epsilon )^2}.
\end{split}
\end{equation*}
The imaginary part can be similarly expressed as a contour integration, and we obtain

\begin{equation}
\begin{split}
2\mathrm{Im}\mathcal{M}_2=\frac{g^2}{36\pi }\int_{1/2-\gamma }^{1/2+\gamma }dx&\frac{e^{2\alpha p^2(1-x)^2}}{4\alpha (1-x)\sqrt{p^2}} \\
&\cdot \frac{e^{-2\alpha A^2}(-8\alpha A^2\sin B+2B\cos B)}{4A^2}, 
\label{eq:imsec}
\end{split}
\end{equation}
where $B=4\alpha (1-x)\sqrt{p^2}A$.

The other terms can be also computed similarly. The term (\ref{eq:unitarythi}) becomes
\begin{equation}
\begin{split}
2\mathrm{Im}\mathcal{M}_3=\frac{g^2}{36\pi }\int_{1/2-\gamma }^{1/2+\gamma }dx&\frac{e^{2\alpha p^2x^2}}{4\alpha x\sqrt{p^2}} \\
&\cdot \frac{e^{-2\alpha A^2}(-8\alpha A^2\sin C+2C\cos C)}{4A^2}, \label{eq:imthi}
\end{split}
\end{equation}
where $C=4\alpha x\sqrt{p^2}A$. The term (\ref{eq:unitaryfor}) becomes
\begin{equation}
\begin{split}
2\mathrm{Im}\mathcal{M}_4=\frac{2g^2e^{\alpha p^2}}{36\pi }\int_{1/2-\gamma }^{1/2+\gamma }dx&\frac{e^{\alpha p^2(1-x)^2}}{2\alpha (1-x)\sqrt{p^2}} \\
&\cdot \frac{e^{-\alpha A^2}(-4\alpha A^2\sin (B/2)+B\cos (B/2))}{4A^2}. \label{eq:imfor}
\end{split}
\end{equation}
The (\ref{eq:unitaryfif}) becomes
\begin{equation}
\begin{split}
2\mathrm{Im}\mathcal{M}_5=\frac{2g^2}{36\pi }\int_{1/2-\gamma }^{1/2+\gamma }dx&\frac{e^{\alpha p^2(2x^2-2x+1)}}{2\alpha (1-2x)\sqrt{p^2}} \\
&\cdot \frac{e^{-2\alpha A^2}(-8\alpha A^2\sin D+2D\cos D)}{4A^2}, \label{eq:imfif}
\end{split}
\end{equation}
where $D=2\alpha (1-2x)\sqrt{p^2}A$.
Finally the term (\ref{eq:unitarysix}) becomes
\begin{equation}
\begin{split}
2\mathrm{Im}\mathcal{M}_6=\frac{2g^2e^{\alpha p^2}}{36\pi }\int_{1/2-\gamma }^{1/2+\gamma }dx&\frac{e^{\alpha p^2x^2}}{2\alpha x\sqrt{p^2}} \\
&\cdot \frac{e^{-\alpha A^2}(-4\alpha A^2\sin (C/2)+C\cos (C/2))}{4A^2}. 
\label{eq:imsix}
\end{split}
\end{equation}

Let us now integrate over $x$ for these results (\ref{eq:imsec})-(\ref{eq:imsix}).
We change the variable $x$ to an angular variable $\varphi$,
\begin{equation}
x=\frac{1}{2}-\gamma \cos \varphi,
\end{equation}
where $0\leq \varphi \leq \pi$. Then $A$ can be expressed as 
\begin{equation}
A=\sqrt{p^2} \,\gamma \sin\varphi.
\end{equation}
The integration can be explicitly carried out 
by using the exponential integral function, which is defined by 
\begin{equation}
\label{eq:defofei}
\mathrm{Ei}(-z)=\int_{\infty}^z \frac{e^{-t}}{t}dt.
\end{equation}
The results are
\begin{align}
2\mathrm{Im}\mathcal{M}_2 &= 
\frac{ig^2e^{2m^2\alpha }}{72\cdot 2\sqrt{p^2}\pi }\left[\mathrm{Ei}(2p^2\alpha (\gamma^2+\gamma e^{-i\varphi }+\gamma^2e^{-2i\varphi }))-(\varphi\rightarrow -\varphi )\right] \bigg|_{0}^{\pi } \label{eq:eifunc1}, \\
2\mathrm{Im}\mathcal{M}_3 &= 
\frac{ig^2e^{2m^2\alpha }}{72\cdot 2\sqrt{p^2}\pi }\left[\mathrm{Ei}(2p^2\alpha (\gamma^2-\gamma e^{-i\varphi }+\gamma^2e^{-2i\varphi }))-(\varphi\rightarrow -\varphi)\right]\bigg|_{0}^{\pi } \label{eq:eifuncsec}, \\
2\mathrm{Im}\mathcal{M}_4 &= 
\frac{2ig^2e^{\alpha (p^2+m^2)}}{72\cdot 2\sqrt{p^2}\pi }\left[\mathrm{Ei}(p^2\alpha (\gamma^2+\gamma e^{-i\varphi }+\gamma^2e^{-2i\varphi }))-(\varphi\rightarrow -\varphi)\right]\bigg|_{0}^{\pi } \label{eq:eifuncthi}, \\
2\mathrm{Im}\mathcal{M}_5 &= 
-\frac{2ig^2e^{2m^2\alpha }}{72\cdot 2\sqrt{p^2}\pi }\left[\mathrm{Ei}(2p^2\alpha (1+e^{2i\varphi }))-(\varphi\rightarrow -\varphi)\right]\bigg|_{0}^{\pi } \label{eq:eifuncfor} ,\\
2\mathrm{Im}\mathcal{M}_6 &= 
\frac{2ig^2e^{\alpha (p^2+m^2)}}{72\cdot 2\sqrt{p^2}\pi }\left[\mathrm{Ei}(p^2\alpha (\gamma^2-\gamma e^{-i\varphi }+\gamma^2e^{-2i\varphi }))-(\varphi\rightarrow -\varphi)\right] \bigg|_{0}^{\pi } \label{eq:eifuncfif}.
\end{align}

\begin{figure}
\begin{center}
\includegraphics[scale=.7]{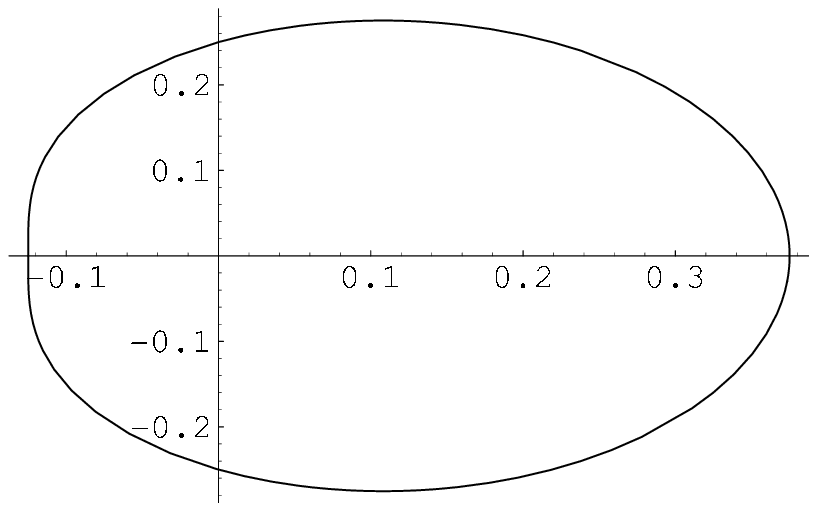}
\includegraphics[scale=.7]{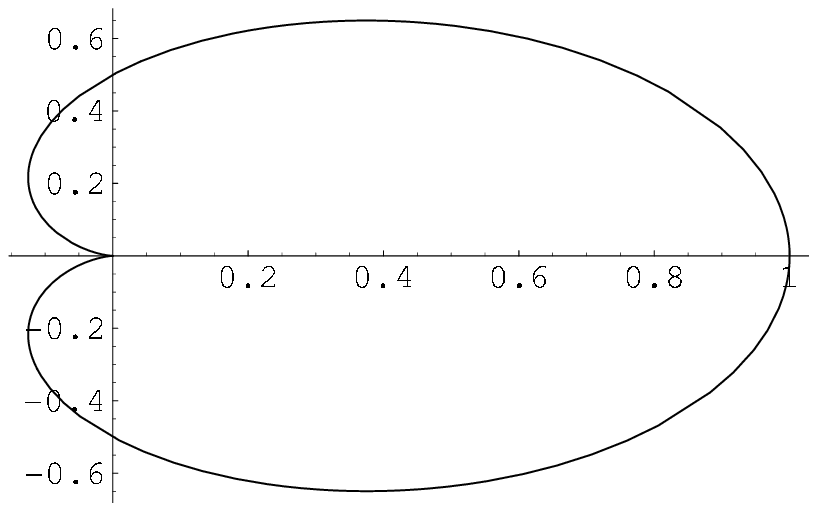}
\end{center}
\caption{The contours of $u$ for $\gamma=\frac14$ (left) and $\gamma=\frac12$ (right).}
\label{fig:contour2}
\end{figure}

Let us first evaluate (\ref{eq:eifunc1}). We first note that
\begin{equation}
\label{eq:pitopi}
\left[\mathrm{Ei}(2p^2\alpha (\gamma^2+\gamma e^{-i\varphi }+\gamma^2e^{-2i\varphi }))-(\varphi\rightarrow-\varphi)
\right] 
\bigg|_{0}^{\pi }= 
\mathrm{Ei}(2p^2\alpha (\gamma^2+\gamma e^{-i\varphi }+\gamma^2e^{-2i\varphi }))\bigg|_{-\pi}^{\pi }.
\end{equation}
Let us consider  
\begin{equation}
u=\gamma^2+\gamma e^{-i\varphi }+\gamma^2e^{-2i\varphi }
\end{equation}
in the argument of the exponential integral function in (\ref{eq:pitopi}). Since $0< \gamma \leq \frac12$,
the trajectory of $u$ in the complex plane for the interval $-\pi \leq \varphi \leq \pi$ is given by 
a closed contour going clockwise around the origin $u=0$, 
as shown in Figure \ref{fig:contour2}. Since the exponential integral
function has a cut stretching between $u=0$ and $u=\infty$, (\ref{eq:pitopi}) can be evaluated from 
the difference of the values of the exponential integral function on distinct sheets. From 
the definition (\ref{eq:defofei}), we find
\begin{equation}
\mathrm{Ei}(2p^2\alpha (\gamma^2+\gamma e^{-i\varphi }+\gamma^2e^{-2i\varphi })) \bigg|_{-\pi}^{\pi } 
=-\oint dt \frac{e^{-t}}{t}= -2 \pi i,
\end{equation}
where the closed contour of $t$ goes counterclockwise around the origin. Thus (\ref{eq:eifunc1}) is evaluated as  
\begin{equation}
2\mathrm{Im}\mathcal{M}_2 = 
\frac{g^2e^{2m^2\alpha }}{72 \sqrt{p^2} }.
\label{eq:m2result}
\end{equation}

We can evaluate (\ref{eq:eifuncsec}), (\ref{eq:eifuncthi}), (\ref{eq:eifuncfif}) in the same manner as above.
The results are 
\begin{align}
2\mathrm{Im}\mathcal{M}_3 &= 
\frac{g^2e^{2m^2\alpha }}{72  \sqrt{p^2} }, \\
2\mathrm{Im}\mathcal{M}_4 &= 
\frac{2g^2e^{\alpha (p^2+m^2)}}{72  \sqrt{p^2} }, \\
2\mathrm{Im}\mathcal{M}_6 &= 
\frac{2g^2e^{\alpha (p^2+m^2)}}{72 \sqrt{p^2} } \label{eq:threeresults}.
\end{align}

\begin{figure}
\begin{center}
\includegraphics[scale=.7]{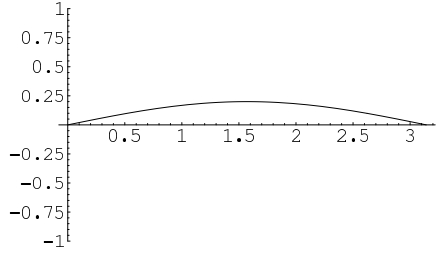}
\end{center}
\caption{The integration contour of $\varphi$ is deformed in the direction of positive imaginary values.}
\label{fig:deform}
\end{figure}

\begin{figure}
\begin{center}
\includegraphics[scale=.7]{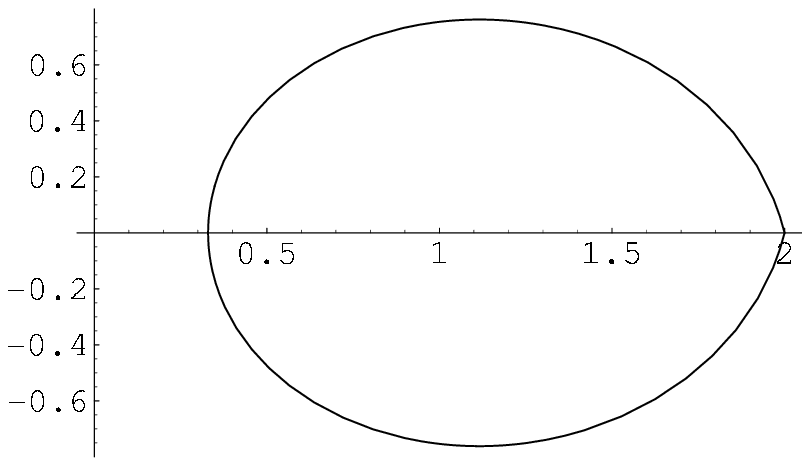}
\includegraphics[scale=.7]{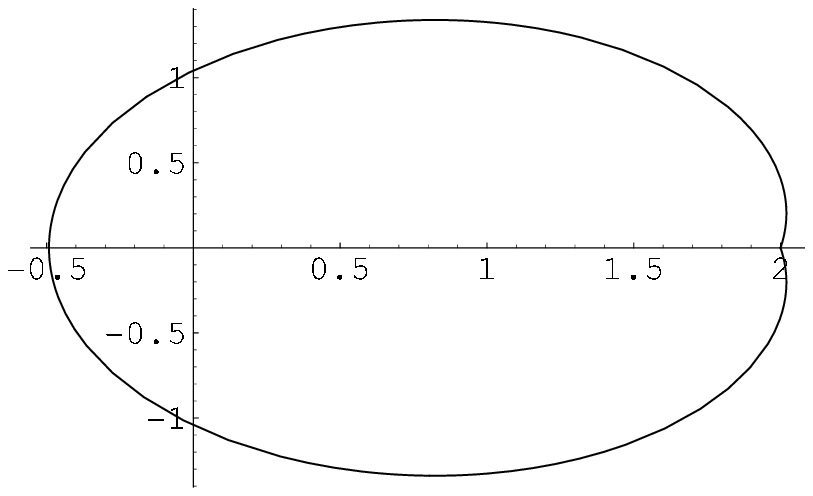}
\end{center}
\caption{The contours of $v_1$ (left) and $v_2$ (right).}
\label{fig:contour3}
\end{figure}

The evaluation of (\ref{eq:eifuncfor}) requires a little care, since the exponential integral function 
is singular at $\varphi =\pi/2$. This is not a physical singularity, since there is no corresponding singularity
in the original integration (\ref{eq:imfif}). In fact, it is allowed to deform the integration contour of $\varphi$ 
apart from  real values. 
Let us deform the path in the direction of positive imaginary values\footnote{The deformation can be in the direction 
of negative imaginary values. This does not change the final result.} as in Figure \ref{fig:deform}. 
Let us consider $v_1=1+e^{2i\varphi}$ and $v_2=1+e^{-2i\varphi}$ in the argument of the first and second terms 
of (\ref{eq:eifuncfor}), respectively. Then $v_1$ does not surround the origin, 
while $v_2$ surrounds the origin clockwise as in Figure \ref{fig:contour3}.  Thus (\ref{eq:eifuncfor}) is evaluated as 
\begin{equation}
2\mathrm{Im}\mathcal{M}_5 = 
\frac{2 g^2e^{2m^2\alpha }}{72 \sqrt{p^2} }.
\label{eq:m5result}
\end{equation}

Collecting all the contributions (\ref{eq:ans1p2}), (\ref{eq:m2result})-(\ref{eq:m5result}), 
we obtain the imaginary part of $\mathcal{M}$ as 
\begin{align}
2\mathrm{Im}\mathcal{M}&=\frac{g^2}{72\sqrt{p^2}}(e^{2\alpha p^2}+4e^{2m^2\alpha }+4e^{\alpha (p^2+m^2)}) \notag \\
&=\frac{g^2}{72\sqrt{p^2}}(e^{\alpha p^2}+2e^{\alpha m^2})^2 .\label{eq:lcut}
\end{align}

On the other hand, the other side of the Cutkosky rule, $\sum |\mathcal{M}|^2$, is given by
\begin{equation}
\begin{split}
\sum |\mathcal{M}|^2=&\frac{1}{2}\bigg(\frac{g}{3}\bigg)^2\int \frac{d^Dq}{(2\pi )^D}\frac{d^Dl}{(2\pi )^D}(e^{\alpha p^2}+e^{\alpha q^2}+e^{\alpha l^2})^2 \\
&\cdot (2\pi )^D\delta ^D(l+q-p) 2\pi \delta (q^2-m^2)2\pi \delta (l^2-m^2). \label{eq:msqured}
\end{split}
\end{equation}
Putting $D=3$ and considering the center-of-mass frame $p=(p,0,0)$, we get
\begin{align}
\sum |\mathcal{M}|^2&=\frac{g^2}{18}\frac{1}{2\pi }\int \frac{d^2q}{4(|\vec{q}|^2+m^2)}\delta (2\sqrt{|\vec{q}|^2+m^2}-p)(e^{\alpha p^2}+2e^{\alpha m^2})^2 \notag \\
&=\frac{g^2}{72\sqrt{p^2}}(e^{\alpha p^2}+2e^{\alpha m^2})^2, \label{eq;finalrhs2}
\end{align}
which agrees with (\ref{eq:lcut}).
Thus we conclude that, in $D=3$, the commutative nonassociative field theory based on 
the product (\ref{eq:p2nonas}) satisfies the one-loop unitarity relation of Figure~\ref{fig:Cutkosky}. 
We also have checked the unitarity in four dimensions in perturbation of $\alpha$, 
and have found that the unitarity is satisfied at least upto the seventh order of $\alpha$.
See the Appendix for details.

\section{Nonassociative $\phi^3$ theory: quartic case} 
\label{sec:p4}
In the previous section we find the one-loop unitarity is satisfied for the commutative 
nonassociative scalar field theory obtained 
from the product defined by (\ref{eq:p2nonas}). But the real part of the one-loop amplitude has an exponential divergence in Minkowski spacetime, which is not
relevant to the one-loop unitarity but may harm the significance of the field theory considered.
In this section, we will define an improved commutative nonassociative product, 
and will study the one-loop unitarity of the scalar field theory obtained from it.  
The price to pay for this change is the loss of exact computation: 
One of the one-loop contributions will be computed in perturbation of a nonassociative parameter $\alpha$. We will
find that the one-loop unitarity holds at least to the order considered.
  
The new product is defined by changing the square of the differentials in (\ref{eq:p2nonas}) into quartic power:
\begin{equation}
\phi(x)\ast \phi(x)=e^{\alpha((\partial_{a}+\partial_{b})^2)^2}\phi(x+a)\phi(x+b)|_{a=b=0}. 
\label{eq:pfornonas}
\end{equation}
In the momentum representation, the new star product is given by 
\begin{equation}
e^{ipx}\ast e^{iqx}=e^{-\alpha \left((p+q)^2\right)^2}e^{i(p+q)x},
\label{eq:momrepquartic}
\end{equation}
where $\alpha>0$.
It is easy to show that this product is commutative but nonassociative. It is obviously Poincar\'{e} invariant. 

In the same way as in Section \ref{sec:p2}, we obtain the following Feynman rules,
\begin{align}
\textrm{propagator} :& ~\frac{i}{p^2-m^2+i\epsilon}, \\
\textrm{three-point vertex} :& ~\frac{-ig}{3}(e^{-\alpha (p^2)^2}+e^{-\alpha (q^2)^2}+e^{-\alpha (k^2)^2}),
\end{align}
where $p, q, k$ are the external momenta.

As in the previous section, we will check the one-loop Cutkosky rule of Figure \ref{fig:Cutkosky} in $D=3$.
The one-loop amplitude is given by
\begin{align}
i\mathcal{M}&=\bigg(\frac{-ig}{3} \bigg)^2\frac{1}{2}\int \frac{d^3q}{(2\pi )^3}
(e^{-\alpha p^4}+e^{-\alpha q^4}+e^{-\alpha (p+q)^4})^2 \notag \\
&\ \ \cdot \frac{i}{q^2-m^2+i\epsilon }\frac{i}{(p+q)^2-m^2+i\epsilon } \notag \\
&=\frac{g^2}{18}\int \frac{d^3q}{(2\pi )^3}\int_0^1dx (e^{-\alpha p^4}+e^{-\alpha q^4}+e^{-\alpha (p+q)^4})^2  \notag \\
&\ \ \cdot \frac{1}{((q+p(1-x))^2+p^2x(1-x)-m^2+i\epsilon)^2},
\label{eq:mrepexp2}
\end{align}
where $p^4$ is the abbreviation for $(p^2)^2$. There occur no exponential divergences, and the integration is 
convergent. After shifting the momentum variable, $q\to q-p(1-x)$, we obtain 
\begin{align}
i\mathcal{M}&=\frac{g^2}{18}\int \frac{d^3q}{(2\pi )^3}\int_0^1dx \left(e^{-\alpha p^4}+e^{-\alpha (q-p(1-x))^4}+e^{-\alpha (q+px)^4}\right)^2  \notag \\
&\cdot \frac{1}{(q^2+p^2x(1-x)-m^2+i\epsilon)^2},
\end{align}
where the exponential factors can be expanded as 
\begin{align}
\big( e^{-\alpha p^4}&+e^{-\alpha (q-p(1-x))^4}+e^{-\alpha (q+px)^4}\big) ^2 \notag \\
&=e^{-2\alpha p^4}+e^{-2\alpha (q-p(1-x))^4}+e^{-2\alpha(q+px)^4} \notag \\
&+2e^{-\alpha p^4-\alpha(q-p(1-x))^4}+2e^{-\alpha (q-p(1-x))^4-\alpha(q+px)^4}+2e^{-\alpha p^4-\alpha(q+px)^4}. \label{eq:expand}
\end{align}
We will compute each contribution in the followings.

Since the first term in (\ref{eq:expand}) does not contain the dependence on $q$, 
we obtain a result similar to (\ref{eq:ans1p2}),
\begin{equation}
2\mathrm{Im}\mathcal{M}_1=\frac{g^2e^{-2\alpha p^4}}{72\sqrt{p^2}}.
\label{eq:fresm1}
\end{equation}

Next let us evaluate the second term in (\ref{eq:expand}). Wick rotating the momenta $p,q$ to Euclidean ones
$p_E,q_E$, and integrating over the angular variables of $q_E$, we get
\begin{equation}
\mathcal{M}_2=\frac{g^2\sqrt{\pi }}{72\sqrt{2\alpha }}\int_0^{\infty}\frac{dq_E}{4\pi ^2}\int_0^1dx~\frac{q_E}{T'}
 \frac{\mathrm{Erf}(\sqrt{2\alpha }(q_E+T')^2)-\mathrm{Erf}(\sqrt{2\alpha }(q_E-T')^2)}{(q_E+A'_{\epsilon})^2(q_E-A'_{\epsilon})^2},
\end{equation}
where $A'_{\epsilon}=\sqrt{-p_E^2x(1-x)-m^2+i\epsilon}$, $T'=p_E(1-x)$ with an abbreviation $p_E=\sqrt{p_E^2}$, 
and the definition of the error function $\mathrm{Erf}(z)$ is given by
\begin{equation}
\mathrm{Erf}(z)=\frac{2}{\sqrt{\pi }}\int_0^ze^{-t^2}dt.
\end{equation} 
The integration over $q_E$ in the evaluation of the imaginary part of $\mathcal{M}_2$ can be carried out
in the same way as in the previous section. The integration can be rewritten as a contour integration around 
the pole, and the result is
\begin{align}
2\mathrm{Im}\mathcal{M}_2=\frac{g^2}{72\cdot 2\pi i}\int _{1/2-\gamma}^{1/2+\gamma}dx&\bigg[\frac{1}{T}(e^{-2\alpha (A+iT)^4}-e^{-2\alpha (A-iT)^4}) \notag \\
&+\frac{i}{A}(e^{-2\alpha (A+iT)^4}+e^{-2\alpha (A-iT)^4})\bigg],
\label{eq:m2four}
\end{align}
where the external momentum has been Wick rotated to the Minkowski one, and $\gamma=\sqrt{(p^2-4 m^2)/4p^2}$, 
$A=\sqrt{p^2x(1-x)-m^2}$, $T=p(1-x)$ with an abbreviation $p= \sqrt{p^2}$. As before, $p^2>4m^2$ or 
$0<\gamma \leq \frac12$ for the imaginary part of the amplitude to exist.  
After the change of variable $x=1/2-\gamma \cos \varphi$, the four terms in (\ref{eq:m2four}) can be collected 
into a simpler expression, 
\begin{equation}
2\mathrm{Im}\mathcal{M}_2=\frac{g^2}{72\cdot 2 p\pi i}\int_0^{2\pi }d\varphi\ e^{-2\alpha (A+iT)^4}\frac{A+iT}{T},
\end{equation}
where $A=p\gamma \sin \varphi$, $T=p/2+p\gamma \cos \varphi$.
Since $A+iT=ip\gamma e^{-i\varphi}+ip/2$, we carry out further the change of variable $z=e^{-i\varphi}$, and 
get
\begin{align}
2\mathrm{Im}\mathcal{M}_2&=\frac{g^2}{72 p \pi i}\oint_C dz e^{-2\alpha (p/2+p\gamma z)^4}\frac{\gamma z +1/2}{z+\gamma z^2+\gamma} \notag \\
&=\frac{g^2}{72 p \pi i}\oint_C dz\frac{e^{-2\alpha (p/2+p\gamma z)^4}(\gamma z+1/2)}{\gamma (z-\bar{\alpha})(z-\bar{\beta})},
\end{align}
where 
\begin{align}
\bar{\alpha}&=\frac{-1+\sqrt{1-4\gamma^2}}{2\gamma}, \\
\bar{\beta}&=\frac{-1-\sqrt{1-4\gamma^2}}{2\gamma},
\end{align}
and $C$ is the counterclockwise circular path with unit radius from the origin.
Since $0 < \gamma \leq 1/2$, $|\bar{\alpha}|<1$ and $|\bar{\beta}|>1$, and $z=\bar{\alpha}$ is the only pole 
in the inside of the unit circle $C$. Evaluating the residue of the pole at $z=\bar{\alpha}$, we get 
\begin{equation}
2\mathrm{Im}\mathcal{M}_2=\frac{g^2}{72 p}e^{-2\alpha m^4}.
\label{eq:fresm2}
\end{equation}

We can easily obtain the contributions from the other terms in (\ref{eq:expand}) except the fifth term.
For example, the third term becomes the same as the second one after the change of variables, $p\to -p$ and $1-x\to x$. The other contributions can be also computed in similar ways. Thus we get 
\begin{align}
2\mathrm{Im}\mathcal{M}_3&=\frac{g^2}{72 p}e^{-2\alpha m^4}, \label{eq:fresm3} \\
2\mathrm{Im}\mathcal{M}_4&=\frac{g^2}{36 p}e^{-\alpha (p^4+m^4)}, \label{eq:fresm4} \\
2\mathrm{Im}\mathcal{M}_6&=\frac{g^2}{36 p}e^{-\alpha (p^4+m^4)}.\label{eq:fresm6} 
\end{align}

Finally we evaluate the contribution from the fifth term in (\ref{eq:expand}).
Wick rotating the momenta $p,q$  to Euclidean ones $p_E,q_E$, 
and carrying out the integration over the angular coordinates of $q_E$, 
we get 
\begin{align}
\mathcal{M}_5&=\frac{g^2\sqrt{\pi }}{36p \alpha^{1/2}}\int_0^{\infty }\frac{q_E dq_E}{4\pi ^2}\int_0^1dxe^{-\alpha\frac{(q_E^2+p_E^2x(1-x))^2}{(1-x)^2+x^2}} \notag \\
&\cdot \frac{1}{(q_E+A'_{\epsilon})^2(q_E-A'_{\epsilon})^2\sqrt{(1-x)^2+x^2}} \notag \\
&\cdot \bigg[\mathrm{Erf}\bigg(\frac{\alpha^{1/2}[(1-x)(q_E+p_E(1-x))^2-x(q_E-p_Ex)^2]}{\sqrt{(1-x)^2+x^2}}\bigg) \notag \\
&-\mathrm{Erf}\bigg(\frac{\alpha^{1/2}[(1-x)(q_E-p_E(1-x))^2-x(q_E+p_Ex)^2]}{\sqrt{(1-x)^2+x^2}}\bigg)\bigg].
\end{align}
The evaluation of the imaginary part and the integration over $q_E$ can be carried out in the same manner as
above, and we get 
\begin{align}
2\mathrm{Im}\mathcal{M}_5&=\frac{g^2 \sqrt{\pi }}{72 \pi i p}\int_{1/2-\gamma}^{1/2+\gamma}dx \frac{1}{A}\bigg[\bigg(\frac{\sqrt{\alpha} m^2 A}{((1-x)^2+x^2)^{3/2}}\bigg)e^{-\frac{\alpha m^4}{(1-x)^2+x^2}} \notag \\
&\cdot \bigg\{\mathrm{Erf}\bigg(\frac{\alpha^{1/2}[(1-x)(A+ip(1-x))^2-x(A-ipx)^2]}{\sqrt{(1-x)^2+x^2}}\bigg) \notag \\
&-\mathrm{Erf}\bigg(\frac{\alpha^{1/2}[(1-x)(A-ip(1-x))^2-x(A+ipx)^2]}{\sqrt{(1-x)^2+x^2}}\bigg)\bigg\} \notag \\
&+e^{-\frac{\alpha m^4}{(1-x)^2+x^2}}\frac{1}{\sqrt{\pi }\sqrt{(1-x)^2+x^2}} \notag \\
&\bigg(\frac{(1-x)(A+ip(1-x))-x(A-ipx)}{\sqrt{(1-x)^2+x^2}}e^{-\frac{\alpha [(1-x)(A+ip(1-x))^2-x(A-ipx)^2]^2}{(1-x)^2+x^2}} \notag \\
&-\frac{(1-x)(A-ip(1-x))-x(A+ipx)}{\sqrt{(1-x)^2+x^2}}e^{-\frac{\alpha [(1-x)(A-ip(1-x))^2-x(A+ipx)^2]^2}{(1-x)^2+x^2}}\bigg)\bigg].
\end{align}
Carrying out the change of variable $x=1/2-\cos{\varphi}$, we obtain a simpler expression,
\begin{align}
2\mathrm{Im}\mathcal{M}_5&=\frac{g^2\sqrt{\pi }}{72\pi i p^2}\int_0^{2\pi }d\varphi \frac{e^{\frac{-\alpha m^4}{1/2+2\gamma^2\cos^2{\varphi}}}}{\sqrt{1/2+2\gamma^2\cos^2{\varphi}}} \notag \\
&\bigg[\frac{\sqrt{\alpha}m^2p\gamma \sin{\varphi}}{1/2+2\gamma^2\cos^2{\varphi}}
\mathrm{Erf}\bigg(\frac{\sqrt{\alpha}(T(A+ipT)^2-S(A-ipS)^2)}{\sqrt{1/2+2\gamma^2\cos^2{\varphi}}}\bigg) \notag \\
&+\frac{T(A+ipT)-S(A-ipS)}{\sqrt{\pi }\sqrt{1/2+2\gamma^2 \cos^2{\varphi}}} \notag \\
&\cdot \exp{\bigg\{-\frac{\alpha(T(A+ipT)^2-S(A-ipS)^2)^2}{1/2+2\gamma^2\cos^2{\varphi}}\bigg\}}\bigg], \label{eq:long}
\end{align}
where
\begin{align}
S&=\frac{1}{2}-\gamma \cos{\varphi}, \\
T&=\frac{1}{2}+\gamma \cos{\varphi}.
\end{align}
We carry out further the change of variable $w=e^{-2i\varphi}$. Then the first term in (\ref{eq:long}) is obtained as
\begin{align}
\frac{\sqrt{2\pi \alpha}g^2 m^2 \gamma}{36p}&\oint_C \frac{dw}{2\pi i}~e^{\frac{-2\alpha m^4 w}{w+\gamma^2 (w+1)^2}}~\frac{w-1}{(w+\gamma^2 (w+1)^2)^{3/2}} \notag \\
&\cdot \mathrm{Erf}\bigg(\frac{-\sqrt{2\alpha}p^2\gamma (\gamma^2w^2+(\gamma^2+5/4)w+1/4)}{\sqrt{w+\gamma^2 (w+1)^2}}\bigg) ,\label{eq:firstterm}
\end{align}
while the second term is obtained as 
\begin{align}
\frac{g^2}{36p}&\oint_C \frac{dw}{2\pi i}\frac{1+2\gamma^2(w+1)}{w+\gamma^2(w+1)^2}
e^{-\frac{\alpha p^4}{8}(1+24\gamma^2 w+16\gamma^4 w^2)}. 
\label{eq:secondterm}
\end{align}
Here the contour $C$ denotes a counterclockwise circular path with unit radius from the origin. 
We have multiplied (\ref{eq:firstterm}) and (\ref{eq:secondterm}) by a factor of $2$, 
because $w$ goes around the origin twice when $\varphi$ varies from $0$ to $2\pi$.

The evaluation of the second contribution (\ref{eq:secondterm}) is straightforward. 
When $\gamma$ is in the range 
$0<\gamma \leq \frac12$, the only pole in the unit circle is at 
\begin{equation}
w=\frac{-2\gamma^2-1+ \sqrt{1+4\gamma^2}}{2 \gamma^2}.
\label{eq:polew}
\end{equation}
Evaluating the residue of the pole, we get
\begin{equation}
\frac{g^2}{36p}e^{-2\alpha m^4}e^{-2\alpha p^2 m^2\big(-1+\sqrt{2-\frac{4m^2}{p^2}}~\big)}.
\label{eq:res1}
\end{equation}

On the other hand, we have not succeeded in evaluating exactly the first contribution (\ref{eq:firstterm}). 
We have computed the contribution in the perturbation of $\alpha$ as follows. 
The error function ${\rm Erf}(z)$ has a series representation, 
\begin{equation}
{\rm Erf}(z)=\frac{2}{\sqrt{\pi}}e^{-z^2} \sum_{k=0}^\infty \frac{2^k z^{2k+1}}{(2k+1)!!}.
\end{equation}
Applying this formula to (\ref{eq:firstterm}), we obtain 
\begin{align}
-\frac{\alpha g^2 \gamma^2m^2 p }{9} \oint_C \frac{dw}{2\pi i} &
\frac{(w-1)(\gamma^2w^2+(\gamma^2+5/4)w+1/4)}{(w+\gamma^2(w+1)^2)^2} 
e^{-\frac{\alpha p^4}{8}(1+24\gamma^2 w+16\gamma^4 w^2)} \nonumber \\
&\cdot
\sum_{k=0}^\infty \frac{1}{(2k+1)!!} 
\left[ \frac{4\alpha p^4 \gamma^2(\gamma^2w^2+(\gamma^2+5/4)w+1/4)^2}{w+\gamma^2(w+1)^2}\right]^k.
\label{eq:expansionm}
\end{align}
In the unit circle $C$, there is only one pole at (\ref{eq:polew}). 
Since the order of the pole becomes higher in
higher order terms of the series, it seems hard to obtain an exact result from the expression (\ref{eq:expansionm}). 
On the other hand, the series in (\ref{eq:expansionm}) can be regarded as a perturbative expansion 
in $\alpha$.
Using Mathematica, we have computed (\ref{eq:expansionm}) in perturbation of $\alpha$ 
upto the seventh order, and have found that (\ref{eq:expansionm}) actually agrees with the expansion of 
\begin{equation}
\frac{g^2}{36p}e^{-2\alpha m^4}\left[1-e^{-2\alpha p^2 m^2\left(-1+\sqrt{2-\frac{4m^2}{p^2}}~\right)}\right].
\label{eq:res2}
\end{equation} 
Thus if we assume this is correct in all orders of $\alpha$, adding (\ref{eq:res1}) and (\ref{eq:res2}),
we finally obtain 
\begin{equation}
2\mathrm{Im}\mathcal{M}_5=\frac{g^2}{36p}e^{-2\alpha m^4}.
\label{eq:fresm5}
\end{equation}

Collecting all the contributions, (\ref{eq:fresm1}), (\ref{eq:fresm2})-(\ref{eq:fresm6}), (\ref{eq:fresm5}), 
the imaginary part of the amplitude is obtained as 
\begin{align}
2\mathrm{Im}\mathcal{M}&=
\frac{g^2}{72p}\left( e^{-2\alpha p^4} +4 e^{-2 \alpha m^4} +4 e^{-2 \alpha (p^4+m^4)} \right) \nonumber \\
&=\frac{g^2}{72p}\left( e^{-\alpha p^4} + 2 e^{-\alpha m^4}\right)^2.
\label{eq:fresleft}
\end{align}

The other side of the Cutkosky rule, $\sum |\mathcal{M}|^2$, can be computed in the similar manner as in 
the preceding section, and we obtain
\begin{equation}
\sum |\mathcal{M}|^2
=\frac{g^2}{72 p }\left(e^{-\alpha p^4}+2 e^{- \alpha m^4}\right)^2,
\end{equation}
which agrees with (\ref{eq:fresleft}). Thus the scalar field theory obtained from the star product 
(\ref{eq:pfornonas}) satisfies the one-loop unitarity of Figure~\ref{fig:Cutkosky}.

\section{Field theories on Poincar\'{e} invariant commutative associative spacetimes} \label{sec:gen}
In this section, we will consider scalar field theories obtained 
from Poincar\'{e} invariant commutative associative algebras.
It will be shown that momentum dependence of couplings does not appear in such cases.
Therefore an algebra must violate commutativity or associativity for scalar field theory 
to have non-trivial properties. This would be a physical restatement of the mathematical
Gelfand-Naimark theorem, which asserts that any associative commutative algebra
is canonically isomorphic to the algebra of functions on some space 
with the usual pointwise multiplication. 

A Poincar\'{e} invariant commutative associative algebra has the general expression, 
\begin{equation}
e^{ip_1x}\ast e^{ip_2x}=f(p_1,p_2)e^{i(p_1+p_2)x},
\end{equation}
where $f(p_1,p_2)$ is a Lorentz invariant function of two momenta $p_1,p_2$ and must satisfy the conditions,
\begin{align}
f(p_1,p_2)&=f(p_2,p_1), \label{eq:com} \\
f(p_1,p_2)f(p_1+p_2,p_3)&=f(p_1,p_2+p_3)f(p_2,p_3), \label{eq:asso}
\end{align}
where (\ref{eq:com}) and (\ref{eq:asso}) are the conditions for commutativity and associativity, respectively.
We also assume a positivity property,
\begin{equation}
f(p,-p)>0
\label{eq:positivity}
\end{equation}
for general $p$. As can be seen below, 
this condition is required for the kinetic term of field theory to be non-singular.

Let us consider scalar $\phi^3$ theory based on the above star product. The action is given by 
\begin{equation}
S=\int d^Dx\bigg[\frac{1}{2}\partial_{\mu}\phi(x)\ast \partial^{\mu}\phi(x)-\frac{1}{2}m^2\phi(x)\ast \phi(x)-\frac{g}{3!}\phi(x)\ast \phi(x)\ast \phi(x)\bigg].
\end{equation}
We define the Fourier transformation of $\phi(x)$ as 
\begin{equation}
\phi(x)=\int \frac{d^Dp}{(2\pi )^D}N(p)\tilde{\phi}(p)e^{-ipx},
\end{equation}
where $N(p)$ is a normalization factor, which will be determined later.
Substituting the Fourier transformation to the action, we get
\begin{align}
S&=\int_p \frac{1}{2}N(p)N(-p)f(p,-p)(p^2-m^2)\tilde{\phi}(p)\tilde{\phi}(-p) \notag \\
&-\frac{g}{3!}\int_p \int_q \int_k N(p)N(q)N(k)f(p,q)f(p+q,k)\delta^D(p+q+k)\tilde{\phi}(p)\tilde{\phi}(q)\tilde{\phi}(k),
\end{align}
where $\int_p$ is $\int d^Dp/(2\pi )^D$.

The physics should not depend on the normalization $N(p)$ of the momentum modes. 
Therefore we can make the propagator of this theory to have the usual form $i/(p^2-m^2+i\epsilon)$ by choosing

\begin{equation}
N(p)=\frac{1}{\sqrt{f(p,-p)}}.
\end{equation}
This choice is allowed from the positivity assumption (\ref{eq:positivity}). 
Then the momentum dependence of the three-point vertex is obtained as 
\begin{equation}
\frac{f(p,q)f(p+q,-p-q)}{\sqrt{f(p,-p)}\sqrt{f(q,-q)}\sqrt{f(-p-q,p+q)}}. \label{eq:3pv}
\end{equation}
For the convenience of the following discussions, let us consider the square of (\ref{eq:3pv}),
\begin{equation}
h=\frac{f^2(p,q)f(p+q,-p-q)}{f(p,-p)f(q,-q)}.
\label{eq:defofh}
\end{equation}
Using the associativity (\ref{eq:asso}),  $f(p,q)f(p+q,-p-q)=f(p,-p)f(q,-p-q)$, 
$h$ can be rewritten as 
\begin{equation}
h=\frac{f(p,q)f(q,-p-q)}{f(q,-q)}. \label{eq:toyu}
\end{equation}
Since $f(p,q)$ is a Lorentz invariant function, it has the property,
\begin{equation}
f(p,-q)=f(-p,q).
\label{eq:changeprop}
\end{equation}
Using (\ref{eq:com}), (\ref{eq:asso}), (\ref{eq:toyu}) and (\ref{eq:changeprop}), we can further show
\begin{equation}
h=f(p,0).
\end{equation}
On the other hand, from (\ref{eq:asso}), we find $f(p,-p)f(0,0)=f(p,-p)f(-p,0)$. Using the positivity 
(\ref{eq:positivity}), we conclude
\begin{equation}
h=f(0,0).
\end{equation}
Therefore $h$ is actually a constant, and the coupling does not have momentum dependence.

The above proof can be generalized to the general coupling $(\ast \phi)^{n+1}$ as follows. The corresponding 
generalization of $h$ is given by
\begin{equation}
h_n=\frac{f\left(\sum_{j=1}^n p_j,-\sum_{j=1}^n p_j\right) \prod_{i=1}^{n-1} f^2\left(\sum_{j=1}^{i} p_j,p_{i+1} \right)}
{\prod_{i=1}^n f\left(p_i,-p_i\right)}.
\end{equation} 
This quantity satisfies an inductive relation,
\begin{equation}
h_n=\frac{f^2(q,p_n) f(q+p_n,-q-p_n)}{f(q,-q)f(p_n,-p_n)}h_{n-1}, 
\end{equation}
where $q=\sum_{i=1}^{n-1} p_i$. The factor in front has actually the same form as (\ref{eq:defofh}). 
Therefore $h_n$ and $h_{n-1}$
are related by a constant multiplication. Since we have shown $h_2$ is a constant, any $h_n$ is also a constant 
by induction.  
Thus we see that commutative associative scalar field theories do not have momentum dependence of couplings.

\section{Discussions and comments} 
\label{sec:discuss}
We find that our commutative nonassociative field theories satisfy the one-loop unitarity of Figure \ref{fig:Cutkosky} in the standard path integral scheme.
This is in contrast with the violation of unitarity in noncommutative field theories with space/time noncommutativity in the standard path integral scheme.
However, our result is based on the explicit computations of the one-loop self-energy diagrams of 
the field theories obtained from certain commutative nonassociative algebras. 
Therefore it is not clear how general our result is, 
i.e. whether unitarity holds for other diagrams and other commutative nonassociative algebras.   
Concerning this question, we point out a qualitative difference in non-locality 
between our field theories and the noncommutative field theories in the following paragraph. 

Noncommutative field theories defined by Moyal product (\ref{eq:moyal}) are non-local in 
real directions. We can easily see this by using the momentum representation as 
\begin{align}
e^{ipx}\ast e^{iqx}&=e^{-\frac{i}{2}p_{\mu}\theta^{\mu\nu}q_{\nu}}e^{i(p+q)x} \nonumber \\
&=e^{ip_{\mu}(x^{\mu}-\frac{1}{4}\theta^{\mu\nu}q_{\nu})}e^{iq_\mu(x^\mu+\frac{1}{4}\theta^{\mu\nu}p_{\nu})}
\label{eq:moyalmom}. 
\end{align}
Therefore when there is space/time noncommutativity, there exists non-locality in time, and field theories will
inevitably show some pathological behaviors such as violation of unitarity \cite{Fujikawa:2004rt}-\cite{Tanaka:1988xk}.
On the other hand, our star product (\ref{eq:momentumrep}) can be rewritten in the form,
\begin{align}
e^{ipx}\ast e^{iqx}&=e^{\alpha (p+q)^2}e^{i(p+q)x} \notag \\
&=e^{ip\cdot (x-i\alpha (p+q))}e^{iq\cdot (x-i\alpha (p+q))}.
\end{align}
One notices that the coordinates are shifted in the imaginary directions.  
This feature is also true for the star product (\ref{eq:momrepquartic}).
Therefore our star products do not have the non-locality in the real time direction, and 
this qualitative difference may 
be the essence of why our field theories do not show the violation of unitarity, 
even though our products contain an infinite number of derivatives with respect to the coordinates.

It is known that noncommutative field theories have the UV-IR mixing property \cite{Filk:1996dm}-\cite{Hayakawa:1999zf}.
The UV-IR mixing is a phenomenon that the ultra-violet divergences appear when external momenta approach zero. 
This occurs because the limit of vanishing external momenta has similar effects as 
the commutative limit $\theta^{\mu\nu} \to 0$ in loop amplitudes. 
On the other hand, the explicit expressions of the one-loop amplitudes (\ref{eq:mrepexp1}), (\ref{eq:mrepexp2})
do not seem to have this sort of similarity between the two limits.  
Therefore our commutative nonassociative field theories will be free from the UV/IR mixing.

The idea of noncommutativity of coordinates stems from the quantization of spacetime. 
But how about nonassociativity? Is there any relation to quantization?
Actually, it seems that nonassociativity can lead to noncommutativity in some situations.
Let us define a right-operation $R(a)$ for an element $a$ of a nonassociative algebra as 
\begin{equation}
R(a)b\equiv a*b.
\end{equation}
Then one finds
\begin{equation}
[R(a),R(b)]c=a*(b*c)-b*(a*c),
\end{equation}
which does not vanish in general even when the nonassociative algebra is commutative. Therefore $R(a)$ is 
noncommutative in general.
It would be interesting to find such effects of noncommutativity in commutative nonassociative field theories.  
 
\section*{Acknowledgments}

Y.S. was supported in part by JSPS Research Fellowships for Young Scientists.
N.S. was supported in part by the Grant-in-Aid for Scientific Research No.13135213, No.16540244 and No.18340061
from the Ministry of Education, Science, Sports and Culture of Japan.

\appendix
\section{Unitarity in four dimensions}
In this section, we check the unitarity in four dimensions in perturbation of $\alpha$.

Let us consider the quadratic case (\ref{eq:p2nonas}). From (\ref{ampfirst}), 
the amplitude is given by
\begin{align*}
\mathcal{M}
&=\frac{g^2}{18}\int \frac{d^4q_E}{(2\pi )^4}\int_0^1 dx \big(e^{-2 \alpha p_E^2} 
+e^{-2\alpha (q_E-p_Ex)^2}+e^{-2\alpha (q_E+p_Ex)^2} \\
&+2e^{-\alpha p_E^2}(e^{-\alpha (q_E-p_Ex)^2}+e^{-\alpha (q_E+p_Ex)^2}) 
+2e^{-\alpha(q_E-p_E(1-x))^2}e^{-\alpha(q_E+p_Ex)^2}\big) \\
&\cdot \frac{1}{(q_E^2+p_E^2x(1-x)+m^2-i\epsilon)^2}. \notag
\end{align*}
Parameterizing the four-momentum $q_E$ with the radial and the spherical coordinates, 
it becomes
\begin{align}
\mathcal{M}&=
\frac{g^2}{18(2\pi )^4}\int_0^{\infty }dq_E q_E^3 \int_0^{2\pi }d\omega \int_0^{\pi }d\varphi \sin{\varphi} \int_0^{\pi}d\theta \sin^2{\theta} \int_0^1 dx \notag \\
&\big(e^{-2 \alpha p_E^2} 
+e^{-2\alpha (q_E^2+p_E^2x^2)}(e^{4\alpha q_Ep_Ex\cos{\theta}}+e^{-4\alpha q_Ep_Ex\cos{\theta}}) \notag \\
&+2e^{-\alpha p_E^2}e^{-\alpha (q_E^2+p_E^2x^2)}(e^{2\alpha q_Ep_Ex\cos{\theta}}+e^{-2\alpha q_Ep_Ex\cos{\theta}}) \notag \\
&+2e^{-\alpha (2q_E^2+p_E^2(2x^2-2x+1)-2p_Eq_E(1-2x)\cos{\theta})}\big) 
\frac{1}{(q_E^2+p_E^2x(1-x)+m^2-i\epsilon)^2} \notag \\
&=\frac{4\pi g^2}{18(2\pi )^4}\int_0^{\infty }dq_E q_E^3 \int_{-1}^{1}dt \sqrt{1-t^2} \int_0^1 dx \notag \\
&(e^{-2 \alpha p_E^2} \label{eq:4dfir} \\
&+2e^{-2\alpha (q_E^2+p_E^2x^2)}\cosh{(4\alpha q_Ep_Ext)} \label{eq:4dsec} \\
&+4e^{-\alpha p_E^2}e^{-\alpha (q_E^2+p_E^2x^2)}\cosh{(2\alpha q_Ep_Ext)} \label{eq:4dthi} \\
&+2e^{-\alpha (2q_E^2+p_E^2(2x^2-2x+1)-2p_Eq_E(1-2x)t)}) \label{eq:4dfor} \\
&\cdot \frac{1}{(q_E^2+p_E^2x(1-x)+m^2-i\epsilon)^2}. \notag
\end{align}

Let us calculate the first term (\ref{eq:4dfir}). After integrating over the variable $t$ and carrying out the replacement $p_E^2=-p^2$, we obtain
\begin{equation}
\mathcal{M}_1=\frac{g^2}{144\pi ^2}e^{2\alpha p^2}\int_0^{\infty }dq_E \int_0^1 dx\frac{q_E^3}{(q_E^2-p^2x(1-x)+m^2-i\epsilon)^2}.
\end{equation}
We can evaluate the imaginary part as we have done in section \ref{sec:p2}. After carrying out the contour integration over $q_E$, we obtain
\begin{align}
2\mathrm{Im}\mathcal{M}_1&=\frac{g^2}{144\pi }e^{2\alpha p^2}\int_{1/2-\gamma}^{1/2+\gamma} dx \notag \\
&=\frac{g^2}{72\pi }e^{2\alpha p^2}\sqrt{\frac{p^2-4m^2}{4p^2}}.
\end{align}
This first order in $\alpha$ is 
\begin{equation}
\label{M11}
2\mathrm{Im}\mathcal{M}_1^{(1)}=\frac{g^2}{36\pi }\alpha p^2\sqrt{\frac{p^2-4m^2}{4p^2}}.
\end{equation}

Next let us calculate the second term (\ref{eq:4dsec}). After integrating over the variable $t$ and carrying out the replacement $p_E^2=-p^2$, we obtain
\begin{align}
\mathcal{M}_2&=\frac{g^2}{36\pi ^2}\int_0^{\infty }dq_E \int_0^1 dx q_E^3 e^{-2\alpha (q_E^2-p^2x^2)}J_1(4\alpha \sqrt{p^2}q_Ex)  \notag \\
&\cdot \frac{1}{4\alpha \sqrt{p^2}q_Ex(q_E+A_{\epsilon})^2(q_E-A_{\epsilon})^2},
\end{align}
where $J_{\nu}(z)$ is the Bessel function.

To obtain the imaginary part, we evaluate
\begin{align}
2\mathrm{Im}\mathcal{M}_2&=\frac{g^2}{36\pi ^2i}\oint dq_E \int_{1/2-\gamma}^{1/2+\gamma} dx q_E^3 e^{-2\alpha (q_E^2-p^2x^2)}J_1(4\alpha \sqrt{p^2}q_Ex) \notag \\
&\cdot \frac{1}{4\alpha \sqrt{p^2}q_Ex(q_E+A+i\epsilon)^2(q_E-A-i\epsilon)^2},
\end{align}
where the contour is given by Figure \ref{fig:contour1}.

After carrying out the contour integration over $q_E$, we obtain
\begin{align}
2\mathrm{Im}\mathcal{M}_2=\frac{g^2}{18\pi } \int_{1/2-\gamma}^{1/2+\gamma} dx \frac{1}{4}e^{-2\alpha (A^2-p^2x^2)}\bigg((1-2\alpha A^2)\frac{J_1(4\alpha \sqrt{p^2}Ax)}{2\alpha \sqrt{p^2}Ax}
-J_2(4\alpha \sqrt{p^2}Ax) \bigg).
\end{align}
Carrying out Taylor expansion in $\alpha $, the first order in $\alpha $ is 
\begin{align}
\label{M21}
2\mathrm{Im}\mathcal{M}_2^{(1)}&=\frac{g^2\alpha }{18\pi } \int_{1/2-\gamma}^{1/2+\gamma} dx\left(m^2+\frac{1}{2}p^2x(-2+3x)\right) \notag \\
&=\frac{g^2\alpha m^2}{18\pi }\sqrt{\frac{p^2-4m^2}{4p^2}}.
\end{align}
In the same way, the third term (\ref{eq:4dthi}) is
\begin{equation}
\label{M31}
2\mathrm{Im}\mathcal{M}_3^{(1)}=\frac{g^2\alpha }{18\pi }(p^2+m^2)\sqrt{\frac{p^2-4m^2}{4p^2}}.
\end{equation}
Finally let us evaluate the forth term (\ref{eq:4dfor}). After integrating over the variable $t$ and carrying out the replacement $p_E^2=-p^2$, we obtain
\begin{align}
\mathcal{M}_4=\frac{g^2}{36\pi ^2}\int_0^{\infty}dq_E\int_0^1 dxq_E^3e^{-\alpha (2q_E^2-p^2(2x^2-2x+1))}\frac{J_1(2\alpha \sqrt{p^2}q_E(1-2x))}{2\alpha \sqrt{p^2}q_E(1-2x)(q_E+A_{\epsilon})^2(q_E-A_{\epsilon})^2}.
\end{align}
To obtain the imaginary part, we evaluate
\begin{align}
2\mathrm{Im}\mathcal{M}_4&=\frac{g^2}{36\pi ^2i}\oint dq_E\int_{1/2-\gamma}^{1/2+\gamma} dxq_E^3e^{-\alpha (2q_E^2-p^2(2x^2-2x+1))} \notag \\
&\cdot \frac{J_1(2\alpha \sqrt{p^2}q_E(1-2x))}{2\alpha \sqrt{p^2}q_E(1-2x)(q_E+A+i\epsilon)^2(q_E-A-i\epsilon)^2}.
\end{align}
After carrying out the contour integration over $q_E$, we obtain
\begin{align}
2\mathrm{Im}\mathcal{M}_4&=\frac{g^2}{18\pi }\int_{1/2-\gamma}^{1/2+\gamma} dx
\frac{1}{4}e^{-\alpha (2A^2+p^2(-1+2x-2x^2))} \notag \\
&\cdot \bigg((1-2\alpha A^2)\frac{J_1(2\alpha \sqrt{p^2}A(1-2x))}{\alpha \sqrt{p^2}A(1-2x)}
-J_2(2\alpha \sqrt{p^2}A(1-2x))\bigg).
\end{align}
Carrying out Taylor expansion in $\alpha $, the first order in $\alpha $ is 
\begin{align}
\label{M41}
2\mathrm{Im}\mathcal{M}_4^{(1)}&=\frac{g^2\alpha }{18\pi }\int_{1/2-\gamma}^{1/2+\gamma} dx \frac{1}{4}(4m^2+p^2(1-6x+6x^2)) \notag \\
&=\frac{g^2\alpha }{18\pi }m^2\sqrt{\frac{p^2-4m^2}{4p^2}}.
\end{align}
Thus, collecting the results (\ref{M11}), (\ref{M21}), (\ref{M31}) and (\ref{M41}), 
the imaginary part of the amplitude in the first order of $\alpha$ is 
\begin{equation}
2\mathrm{Im}\mathcal{M}=\frac{\alpha g^2}{12\pi }(p^2+2m^2)\sqrt{\frac{p^2-4m^2}{4p^2}}.
\end{equation}

On the other hand, the other side of the Cutkosky rule, $\Sigma |\mathcal{M}|^2$ is given by
\begin{align}
\Sigma |\mathcal{M}|^2&=\frac{1}{2}\left(\frac{g}{3}\right)^2 \int \frac{d^4q}{(2\pi )^4}\frac{d^4l}{(2\pi )^4}(e^{\alpha p^2}+e^{\alpha q^2}+e^{\alpha l^2})^2 \notag \\
&\cdot (2\pi )^4 \delta^4(l+q-p)2\pi \delta(q^2-m^2)2\pi \delta(l^2-m^2) \notag \\
&=\frac{g^2}{18(2\pi )^2}(e^{\alpha p^2}+2e^{\alpha m^2})^2\int\frac{d^3q}{2\sqrt{\vec{q}^2+m^2}}\int\frac{d^3l}{2\sqrt{\vec{l}^2+m^2}}\delta^4(l+q-p) \notag \\
&=\frac{g^2}{72\pi}\sqrt{\frac{p^2-4m^2}{4p^2}}(e^{\alpha p^2}+2e^{\alpha m^2})^2.
\end{align}
The first order of $\alpha$ of $\Sigma |\mathcal{M}|^2$ is
\begin{equation}
\Sigma |\mathcal{M}|^{2(1)}=\frac{g^2\alpha}{12\pi}\sqrt{\frac{p^2-4m^2}{4p^2}}(p^2+2m^2).
\end{equation}
Thus we find that the unitarity is satisfied in the first order of $\alpha$ 
in four dimensions.
We can check the unitarity in higher orders of $\alpha $ in the same way, and 
have found that the unitarity is satisfied at least up to the seventh order of $\alpha$,
using the Mathematica.

\end{document}